\documentclass[prb,twocolumn,english,amsmath,amssymb,floatfix,superscriptaddress,longbibliography,preprintnumbers]{revtex4-2}

\usepackage{chemformula} % Formula subscripts using \ch{}
\usepackage[T1]{fontenc} % Use modern font encodings
\usepackage{mathptmx}
\begin{document}

\title{Micrometer – resolution fluorescence and lifetime mappings  of CsPbBr$_3$ nanocrystal films coupled with a TiO$_2$ grating}

\author{Viet Anh Nguyen}
\affiliation{Center of Environmental Intelligence and Design, College of Engineering and Computer Science, VinUniversity, Gia Lam district, Hanoi 14000, Vietnam}

\author{Linh Thi Dieu Nguyen}
\affiliation{Center of Environmental Intelligence and Design, College of Engineering and Computer Science, VinUniversity, Gia Lam district, Hanoi 14000, Vietnam}

\author{Thi Thu Ha Do}
\affiliation{Institute of Materials Research and Engineering, Agency for Science Technology and Research (A*STAR), 2 Fusionopolis Way, 138634, Singapore}

\author{Ye Wu}
\affiliation{Department of Materials Science and Engineering, and Centre for Functional Photonics (CFP), City University of Hong Kong, 83 Tat Chee Avenue, Hong Kong SAR, 999077 P. R. China}

\author{Aleksandr A. Sergeev}
\affiliation{Department of  Physics, Hong Kong University of Science and Technology, Clear Water Bay, Kowloon, Hong Kong, P. R. China}

\author{Ding Zhu}
\affiliation{Department of Materials Science and Engineering, and Centre for Functional Photonics (CFP), City University of Hong Kong, 83 Tat Chee Avenue, Hong Kong SAR, 999077 P. R. China}

\author{Vytautas Valuckas}
\affiliation{Department of  Physics, Hong Kong University of Science and Technology, Clear Water Bay, Kowloon, Hong Kong, P. R. China}

\author{Duong Pham}
\affiliation{Institute of Physics, Vietnam Academy of Science and Technology, 10 Dao Tan Street, Hanoi, Vietnam}

\author{Hai Xuan Son Bui}
\affiliation{Center of Environmental Intelligence and Design, College of Engineering and Computer Science, VinUniversity, Gia Lam district, Hanoi 14000, Vietnam}
\affiliation{Graduate University of Science and Technology, 18 Hoang Quoc Viet Street, Hanoi, Vietnam}

\author{Duy Mai Hoang}
\affiliation{College of Health Science, VinUniversity, Gia Lam district, Hanoi 14000, Vietnam}

\author{Son Tung Bui}
\affiliation{Graduate University of Science and Technology, 18 Hoang Quoc Viet Street, Hanoi, Vietnam}

\author{Xuan Khuyen Bui}
\affiliation{Institute of Materials Science, Vietnam Academy of Science and Technology, 18 Hoang Quoc Viet Street, Hanoi, Vietnam}

\author{Binh Thanh Nguyen}
\affiliation{Institute of Physics, Vietnam Academy of Science and Technology, 10 Dao Tan Street, Hanoi, Vietnam}

\author{Hai Son Nguyen}
\affiliation{Univ Lyon, Ecole Centrale de Lyon, CNRS, INSA Lyon, Université Claude Bernard Lyon 1, CPE Lyon, CNRS, INL, UMR5270, Ecully 69130, France}
\affiliation{Institut Universitaire de France (IUF), 75231 Paris, France}

\author{Lam Dinh Vu}
\affiliation{Graduate University of Science and Technology, 18 Hoang Quoc Viet Street, Hanoi, Vietnam}

\author{Andrey Rogach}
\affiliation{Department of Materials Science and Engineering, and Centre for Functional Photonics (CFP), City University of Hong Kong, 83 Tat Chee Avenue, Hong Kong SAR, 999077 P. R. China}

\author{Son Tung Ha}
%\email{ha_son_tung@imre.a-star.edu.sg}
\affiliation{Institute of Materials Research and Engineering, Agency for Science Technology and Research (A*STAR), 2 Fusionopolis Way, 138634, Singapore}

\author{Quynh Le-Van}\email{quynh.lv@vinuni.edu.vn}
\affiliation{Center of Environmental Intelligence and Design, College of Engineering and Computer Science, VinUniversity, Gia Lam district, Hanoi 14000, Vietnam}

\begin{abstract}

Enhancing light emission from perovskite nanocrystal (NC) films is essential in light-emitting devices, as their conventional stacks often restrict the escape of emitted light. This work addresses this challenge by employing a TiO$_2$ grating to enhance light extraction and shape the emission of CsPbBr$_3$ nanocrystal films. Angle-resolved photoluminescence (PL) demonstrated a tenfold increase in emission intensity by coupling the Bloch resonances of the grating with the spontaneous emission of the perovskite NCs. Fluorescence lifetime imaging microscopy (FLIM) provided micrometer-resolution mapping of both PL intensity and lifetime across a large area, revealing a decrease in PL lifetime from 8.2 ns for NC films on glass to 6.1 ns on the TiO$_2$ grating. Back focal plane (BFP) spectroscopy confirmed how the Bloch resonances transformed the unpolarized, spatially incoherent emission of NCs into polarized and directed light. These findings provide further insights into the interactions between dielectric nanostructures and perovskite NC films, offering possible pathways for designing better performing perovskite optoelectronic devices.\\
\onecolumngrid
\centerline{\includegraphics[width=0.4\textwidth]{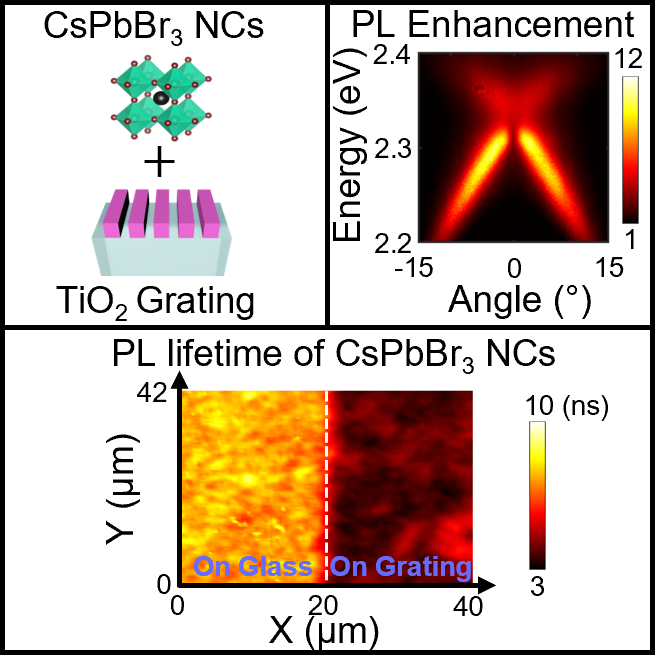} }
\end{abstract}

\maketitle

Metal halide perovskites have sparked considerable interest in the development of new generations of photovoltaics,\cite{Swarnkar2016,Hu2021,Ma2024} light-emitting diodes (LED),\cite{Chen2017,Yan2018,Chiba2018,Lin2018,Kim2021,Chen2021,Wan2023} photodetectors\cite{Ramasamy2016, Wang2021,Zhang2024} and lasers.\cite{Xing2014,Zhang2014,Zhu2015,Zhang2016,Tiguntseva2020,Zhang2021,AhmadKamal2023,Xing2024} All-inorganic halide perovskites, specifically those in the CsPbBr$_3$ family (where X represents halides such as Cl, Br, or I), exhibit advantageous properties such as thermal stability and defect tolerance.\cite{Huang2016,Shamsi2019,Yuan2020,Dey2021} A critical aspect of Cs-based colloidal perovskite NCs is their light-emission capability, which is an essential metric for assessing the LED’s performance. The external quantum efficiency (EQE) of a LED is defined as \(\eta_{ex}=\eta_{in} \cdot \eta_{op}\), where \(\eta_{in}\) represents the internal quantum efficiency, which is the ratio of internally generated photons to the total carriers injected into the active layer, and \(\eta_{op}\) is the optical efficiency, which describes the extraction of light out of the device. Extensive studies have been carried out to obtain perovskite NCs with a high PL quantum yield (PLQY), which is directly related to \(\eta_{in}\). Although perovskite NCs have demonstrated near-unity PLQY in solution, their internal efficiency in solid films often decreases significantly, dropping to as low as 2$\%$ in some reported cases due to non-radiative transfer between densely packed NCs.\cite{Dai2018,Kim2015} Several strategies have been proposed to increase the PLQY of the perovskite NC films to values as high as 70$\%$, including ligand engineering\cite{Ban2018} and the use of sub-micrometer structures.\cite{Cao2018} Direct precursor deposition has yielded even higher PLQY, exceeding 85$\%$.\cite{Zeng2021} Kumar et al reported an increased PLQY of the NC solids approaching nearly 100$\%$ compared to 82$\%$ in solution, attributed to the aggregation-induced emission.\cite{Kumar2019}

Despite the extensive efforts focused on improving the quality and the PLQY of perovskite films, the reported EQE of perovskite NCs-based LEDs remains inferior to their PLQY values, typically around 20$\%$.\cite{Dong2020,Kumar2022,Li2024} As the field of all-inorganic halide perovskite NCs rapidly evolves, addressing the challenges associated with optimizing the EQE of the respective LEDs remains a critical focus. A major reason limiting high EQE values is that the emitted light is often trapped in waveguides formed by vertical confinement within the stacks, or by the low value of \(\eta_{op}\). Several approaches have been implemented to address the issue of the large optical contrast between the active and injection layers of such LEDs, including patterned metasurfaces\cite{Dang2020,Dang2022,wang_directional_2023,lu_engineering_2020,Ali2024,Marcato2024} and superlattices of perovskite NCs.\cite{Kumar2022,Cherniukh2021} Alongside self-patterned perovskites, nanophotonic structures, such as plasmonic or dielectric gratings, have been extensively utilized to tailor key aspects of fluorophore emission - such as brightness, directivity, and polarization - by altering the local density of states, a phenomenon known as Purcell effect.\cite{Curto2010,Hoang2015,Krasnok2015,He2024} While this approach has been achieved in various ensembles of fluorophores, including organic dyes,\cite{Berghuis2019,Daskalakis2019,Liu2019,Berghuis2020,Heilmann2020,Berghuis2022,S2023,Bhaskar2024} quantum wells,\cite{Benisty2008,Ndiaye2022,Abdelkhalik2023} semiconductor NCs,\cite{LeVan2016,Coste2017,Wang2018,Yadav2021,Xiong2022,monin_controlling_2023,Ha2023,Sergeeva2023,Bossavit,bailly_2d_2024} and 2D materials,\cite{Wang2016,Duong2018,Wang2019,Turunen2022} the use of resonant 
nanostructures to control spontaneous emission of the perovskite NC film has been less explored.

In this study, we report improved light extraction from the perovskite CsPbBr$_3$ NCs films achieved by coupling them with a low-loss dielectric titanium dioxide (TiO$_2$) grating. The grating supports surface waves, referred to Bloch (guided) resonances, which facilitate the dispersion folding into the radiated light cone. The NC films were prepared using a standard spin-coating technique commonly employed in the fabrication of perovskite-based optoelectronics. The grating was designed to enhance the PL emission of the NC films and to simultaneously improve the light extraction of the film. BFP spectroscopy and FLIM measurements were employed to investigate the enhancements both in momentum and real space for the samples where CsPbBr$_3$ NCs films were deposited on top of the TiO$_2$ grating as compared to those on glass substrates. We obtained the spatial distribution of PL enhancement and PL lifetime at the micrometer resolution. Numerical calculations confirmed that the optical field enhancements occurred in the perovskite NCs layer when deposited on the TiO$_2$ grating, facilitating the radiative processes in the NCs via the Purcell effect. While the light is confined in the perovskite NC film deposited on glass, it can be extracted into free space by coupling with the Bloch resonances existing in perovskite NCs deposited on TiO$_2$ gratings. BFP spectroscopy for individual frequencies revealed the coupling of the NCs’ emission with Bloch resonances which accounted for the improvement of light extraction. 
We started this work by studying thin films of CsPbBr$_3$ NCs on a glass substrate. CsPbBr$_3$ NCs were synthesized by a hot injection method;\cite{Vighnesh2022,Li2024a} details of the synthesis are described in Section 1 of the supporting information (SI). CsPbBr$_3$ NCs were washed multiple times by centrifugation and dissolved in octane with a concentration of 10 mg/ml. A thin (400 nm) layer of these NCs was obtained by spin-coating on a glass substrate and baked at 80 $^{\circ}$C for 5 min. Figure 1a provides a sketch of a CsPbBr$_3$ NCs film deposited on a glass substrate. The transmission electron microscopy (TEM) image of the NCs and the X-ray diffraction (XRD) pattern of their film are shown in Figs. 1b and 1c, respectively. The TEM image indicates the average side 12.2 ± 0.5 nm for the nanocube-shaped crystals, while the XRD pattern aligns with the reference pattern of the CsPbBr$_3$ NCs (PDF$\#$54-0752). The pattern exhibits characteristic peaks at 2$\theta$ values of 15.1$^{\circ}$, 21.4$^{\circ}$, 30.5$^{\circ}$, 33.9$^{\circ}$, 37.7$^{\circ}$, and 43.5$^{\circ}$, corresponding to the (100), (110), (200), (210), (211), and (202) crystallographic planes of the cubic phase structure of CsPbBr$_3$ NCs. The absorption and PL spectra of the perovskite NCs film are shown in Fig. 1d. A well-defined excitonic peak in the absorption spectrum is observed at 2.42 eV (512 nm), while PL peaks at 2.39 eV (520 nm) with a linewidth of 68 meV when excited by a continuous laser ($\lambda_{ex}$ = 405 nm). It is noted that the   NCs film still exhibit a decent PL even after a week of exposure to humid air. The PLQY of the NCs film measured after such a long exposure to air was found to be 5.6 ± 0.7 $\%$, as determined by three-measurement technique using an integrated sphere in a spectrophotometer. Details of the PLQY measurement are provided in Fig. S1 of the SI. 

\begin{figure}[ht]
\centering
\includegraphics[width=1\linewidth]{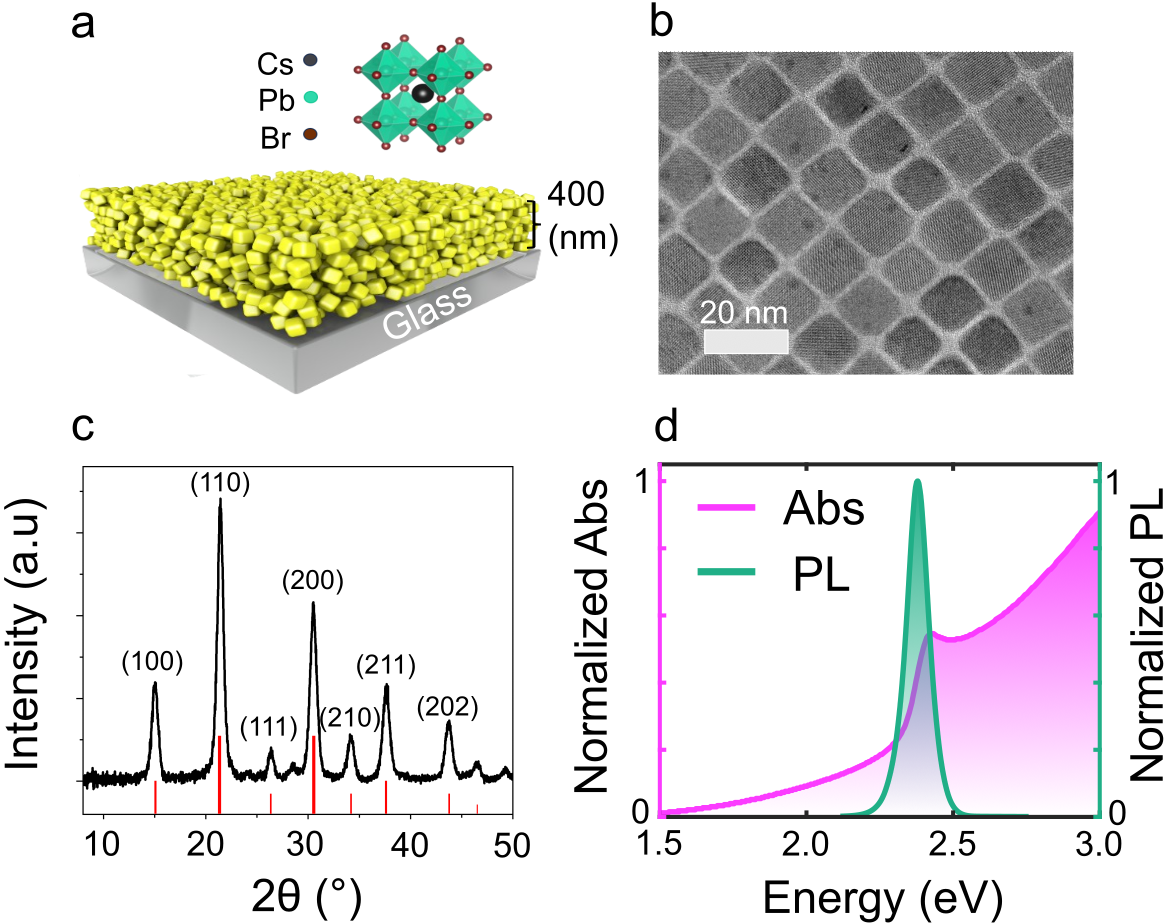}
\caption{\textbf{Characterization of the perovskite NCs film on glass}. a) Sketch of a layer of perovskite NCs film (thickness = 400 nm) on a glass substrate. b) TEM image of CsPbBr$_3$ NCs used in this work. c) XRD pattern of the CsPbBr$_3$ NC film (in black) compared with a standard Inorganic Crystal Structure Database spectrum of its cubic phase (in red). d) Absorption and PL spectra of the perovskite NC film.}
\label{fig1}
\end{figure}

Next, we present 1D TiO$_2$ grating that supports an optical mode overlapping with the PL emission of the perovskite NCs film. Figure 2a illustrates the complete system, which consists of TiO$_2$ grating on a glass substrate, covered by a layer of CsPbBr$_3$ NC film. The TiO$_2$ grating was fabricated firstly by depositing a continuous, 120 nm thick, TiO$_2$ layer on the glass substrate via the sputtering process. The TiO$_2$ film was subsequently patterned by electron beam lithography followed by dry and wet etching processes. Full details of the TiO$_2$ grating fabrication are presented in Section 2 of the SI. The scanning electron microscopy (SEM) image of the final structure is shown in Fig. 2b, featuring a period of p = 260 nm and a width of w = 158 nm. Probing the Bloch resonances is a challenging task due to the stacked vertical structure of the sample, which causes them to be hidden within the scattering spectra. To reveal the Bloch resonances of the grating covered by the perovskite NCs film, we employed a cross-polarization angle-resolved reflectivity measurements.\cite{Nguyen2023} A system comprising of linear polarizer (LP) and half-wave plate (HWP) was utilized to select diagonally polarized ($\text{$\nearrow$\llap{$\swarrow$}}$) light from an unpolarized light source, which impinged upon the structure at 45$^{\circ}$ with respect to the x-axis. A second set of the LP and HWP was configured to analyze the antidiagonal polarized light ($\text{$\nwarrow$\llap{$\searrow$}}$), oriented at – 45$^{\circ}$ with respect to the x-axis, directed toward the spectrometer. Cross polarization reflectivity eliminated the background reflections, allowing us to probe the pure guided resonances in the slap of grating-perovskite structure. Details of the measurements are presented in Section 3 of the SI. The high-intensity curves in the dispersion graph presented in Fig. 2c correspond to Bloch resonances, which simultaneously appear in transverse electric (TE) and transverse magnetic (TM) polarizations. This occurs due to the excitation conditions of the cross-polarization measurements, which stimulate both modes. In our experimental configuration, the TE mode is defined as having the electric field aligned along the groove direction (the y-axis), while the TM mode has the electric field polarized along the x-axis. The TE mode exhibits two branches: one with lower energy that vanishes at normal incident, referred to dark mode, and another at higher   energy that displays a faint line with a smaller slope compared to the former, known as bright mode. The bright mode in both TE and TM configurations does not pass through 2.42 eV due to the absorption of excitons in the perovskite NCs layer. Here we demonstrate that the interactions between perovskite NCs and the TiO$_2$ grating are similar for both TE and TM modes; however, the coupling efficiency is slightly higher for the former. Therefore, for simplicity, we will focus our analysis on the TE mode. We validated the experimental results by performing numerical simulation of the angle- resolved reflectivity for the studied system with Comsol Multiphysics. Figure 2d presents the simulated angle-resolved reflectance of the TiO$_2$ grating coated by the perovskite NC film. The simulation was conducted by applying Floquet boundary conditions for an infinite grating lattice.  Details of the simulation model are provided in Section 4 of the SI. Figure 2e shows the side view  of the electric field modulus distributed within the perovskite NC film for the three cases: (i) on the  glass substrate (left panel), (ii) on the TiO$_2$ grating at point A marked in Fig. 2d (middle panel), and (iii) on the TiO$_2$ grating at point B marked in Fig. 2d (right panel). While the electric field is nearly absent in case (i) and slightly improved in case (ii), it becomes significantly enhanced in the last case (iii). The electric field is amplified within a large volume fraction of the NC layer, not only in the close vicinity of the grating but also far into the perovskite layer. 

\begin{figure}[ht]
\centering
\includegraphics[width=1\linewidth]{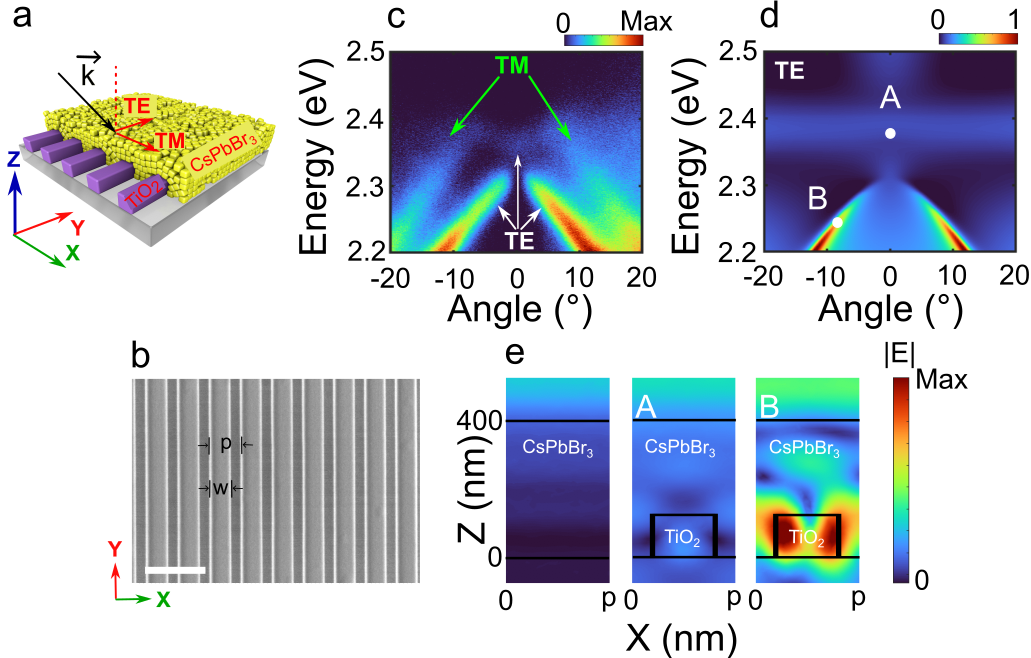}
\caption{\textbf{Characteristics of TiO$_2$ grating covered by a CsPbBr$_3$ NC film}. a) A sketch of the system consists of a layer of 1D TiO$_2$ grating on a glass substrate covered by a film of CsPbBr$_3$ NCs. The lower inset shows the coordinates of the samples with a guided mode that exists in the hybrid layer of grating and perovskite film. b) SEM image of the TiO$_2$ grating with a period p = 260 nm and a width w = 158 nm. The scale bar is 500 nm. c) Measured angle-resolved reflectivity using cross-polarization. Two resonant modes (TE and TM) are displayed. d) Simulated angle-resolved reflectivity of the structure in TE mode performed with Comsol Multiphysics. e) The electric field distribution of two selected points A and B in Fig. 2e. Point A is at 2.39 eV and 0$^{\circ}$ and its reference point of the perovskite NC film on the glass substrate (left panel); Point B is on one of the Bloch resonances at 2.25 eV and 7.5$^{\circ}$. The electric field is concentrated and enhanced in the grating structure in point A, whereas the majority of the field is enhanced in the perovskite layer in point B.}
\label{fig2}
\end{figure}

We now investigate how the PL of the system behaves when the perovskite NC film is placed on glass substrate or TiO$_2$ grating. Figures 3a and 3b display the angle-resolved PL maps for a perovskite film on the glass substrate and on TiO$_2$ grating, respectively. The spectra were measured under identical excitation conditions (using a 405 nm continuous laser with low pumping power (0.4 $\mu$W/$\mu$m$^{2}$) to ensure linear response upon excitation intensity, see Fig. S1b for pumping curve. Here, the excitation enhancement is considered negligible. Further discussion about the excitation is provided in Section 4 and Fig. S4 of the SI. The integration time was maintained  at 0.1 s and the slit width was kept at 30 $\mu$m. The emitted signal from the objective’s back-focal plane was directed onto the entrance slit of the monochromator and recorded by a spectral camera. Details of the angle-resolve PL measurements are described in Section 3 of the SI. Figure 3a shows an angle-resolved PL map of the perovskite NC film on a glass substrate measured in TE mode. The PL intensity exhibits a peak at 2.37 eV (523 nm) and is uniform across the angular space captured by the objective. This uniformity reflects the spatially incoherent characteristics of the perovskite NCs’ emission. The same signal is observed when measured in TM mode, indicating that the PL intensity of the perovskite NCs is unpolarized. However, when we switched to the perovskite NC film on the TiO$_2$ grating, we observed a strongly modified PL profile with distinct features compared to the NC film on the glass substrate. Figure 3b shows the  angle-resolved PL map of the perovskite NC film on TiO$_2$ grating in TE mode. The PL intensity increases at the main emission peak of the NC film and two new lobes appear, vanishing at 0$^{\circ}$ ($\Gamma$ point), with intensity declining at increased angles. The new features observed in the PL map are consistent with the reflectance map shown in Fig. 2c. A similar effect with different lobes was also observed when the NCs’ emission coupled with the TM mode of the grating, as illustrated in Fig. S5a of the SI. To elucidate the spectral shaping when the NC film’s emission coupled with the Bloch resonances of the grating, spectra of the PL emission on the glass selected at 5$^{\circ}$ and on the grating at 2$^{\circ}$, 5$^{\circ}$, and 8$^{\circ}$ are presented in Fig. 3c. For the latter case, the PL signal exhibits a pronounced enhancement across all emission spectrum when compared to that of the NC film on glass. This enhancement is particularly significant at the peak emission wavelength of the NCs (2.38 eV) and exhibits a slight redshift as increased angles. In addition, a secondary peak was observed at a lower energy of 2.31  eV (2$^{\circ}$), 2.29 V (5$^{\circ}$), and 2.25 eV (8$^{\circ}$). The emerged peaks are attributable to the improved light outcoupling facilitated by the interaction between the Bloch resonances of the TiO$_2$ grating and the emission from the perovskite NCs. Such behavior is consistent with the observations of electric field distribution illustrated in Fig. 2e, where the strongest localization of electromagnetic field is located in the vicinity of TiO$_2$ grating. To provide insight into the underlying interaction mechanism, a schematic representation in Fig. 3d depicts a star-point that symbolizes an emitter located within the NC film layer on both the glass substrate and TiO$_2$ grating. The emission signal in the case of the glass substrate is governed by the basic laws of refraction. The effective refractive index of the perovskite NC film is calculated to be 1.69 and corresponds to a light escape cone angle of 36.3$^{\circ}$. The introduction of TiO$_2$ gratings enhances the escape of light constrained within the perovskite NCs film. This enhancement is a product of the outcoupling processes between emitted light and photonic modes associated with the grating. The normalized spatial PL intensity of the examined area, collected on a FLIM setup, is illustrated in Fig. 3e. A notable contrast is clearly noted between PL intensities of the NC film on glass and the TiO$_2$ grating, with the latter being approximately three times stronger. This contrast in PL intensity and lifetime is contrary to the expected behavior of fluorescent probes experiencing self-quenching at high concentrations under the electrical excitation conditions.\cite{Meredith2023} The PL lifetime mapping extracted from single exponential fitting of PL data from Fig. 3e is illustrated in Fig. 3f; the quality of the fitting is presented in Fig. S6 of the SI. One can see that PL lifetime of the perovskite NCs on the TiO$_2$ grating noticeably decreases compared to that on the glass. A similar result was also observed in TM mode, as shown in Fig. S7a-b of the SI. Figure 3g displays the extracted PL intensity and lifetime values along the dotted line shown in Figs. 3e and 3f. A profound contrast occurs at the edge of the grating area, where a reduction in PL lifetime goes ahead with increase in PL intensity as the perovskite NCs interact with the TiO$_2$ grating. Specifically, the average PL lifetime of the perovskite NCs on the TiO$_2$ grating reduced to 5.9 ns from 8.16 ns on the glass, while the normalized PL intensity increased to 0.71 on the grating from 0.27 on the glass. The PL intensity enhancement and lifetime reduction are slightly disproportionate, which could be explained by the dominance of zero-phonon emission rate occurring in low PLQY emitters.\cite{Lefaucher2024,SonBui24} 

\begin{figure}
\centering
\includegraphics[width=1\linewidth]{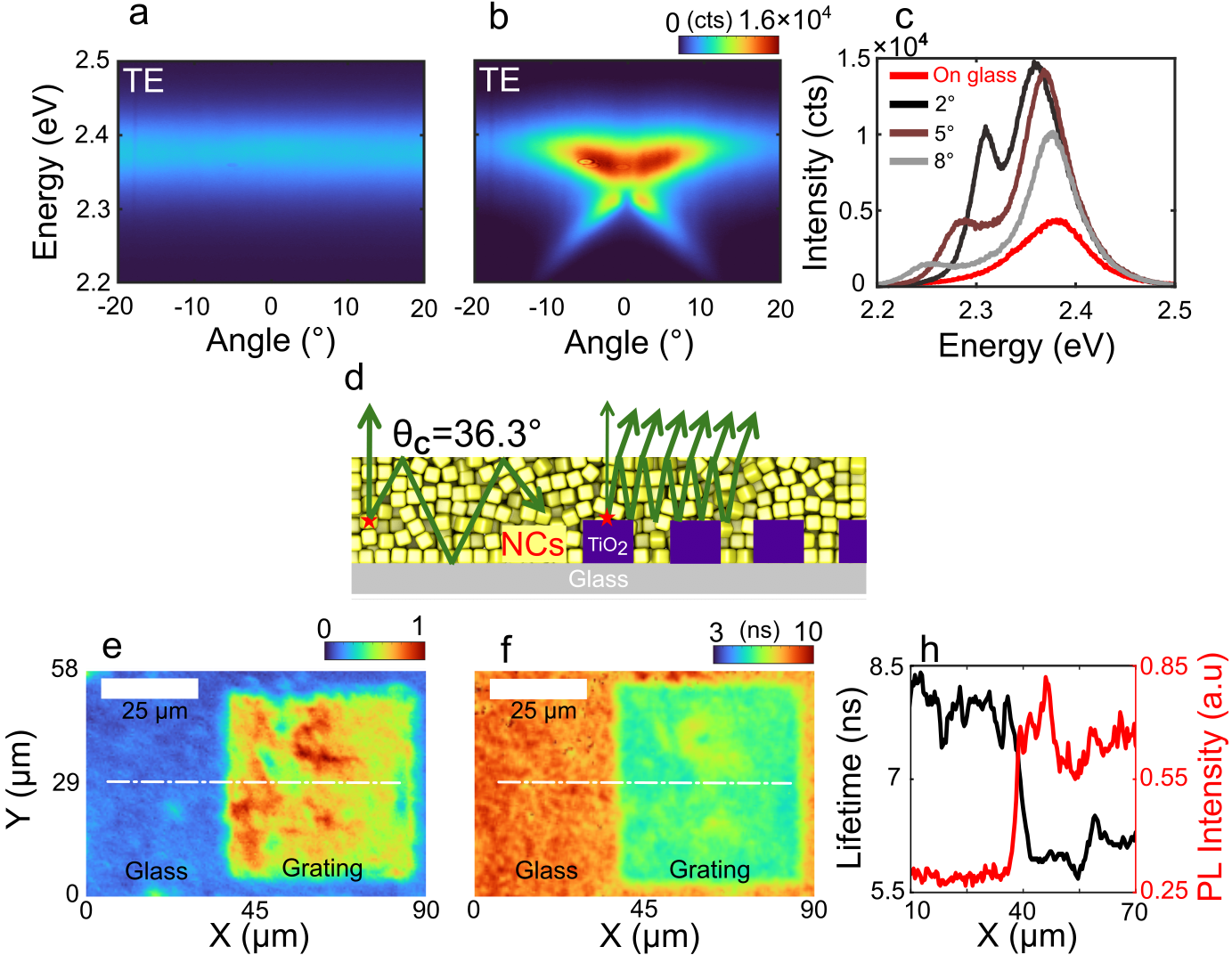}
\caption{\textbf{PL emission of CsPbBr$_3$ NC film on glass and on TiO$_2$ grating}. PL emission of CsPbBr$_3$ NC film on glass and on TiO$_2$ grating. Angle-resolved PL spectra of CsPbBr$_3$ NCs on a) glass substrate and b) on TiO$_2$ grating. c) The spectra of the CsPbBr$_3$ NC film collected on glass at 5$^{\circ}$ (red), and on the TiO$_2$ grating at 2$^{\circ}$ (black), 5$^{\circ}$ ( brown) and 8$^{\circ}$ (grey). d) A sketch illustrating the mechanism of enhancement of emitter light. e) PL intensity distribution of the CsPbBr$_3$ NCs film on the glass substrate and on TiO$_2$ grating with the scanned area 58 $\mu$m x 90 $\mu$m in FLIM measurement. f) PL lifetime extracted from PL map in e). The white dash lines in e) and f) are horizontal cuts passing through the perovskite NCs on glass and on the TiO$_2$ grating. g) The PL lifetime and normalized PL intensity as a function of the position along the white dashed line. The PL intensity (lifetime) of CsPbBr$_3$ NCs on glass and on the TiO$_2$ grating is 0.27 (8.16 ns) and 0.71 (5.9 ns), respectively.}
\label{fig3}
\end{figure}

\begin{figure}
\centering
\includegraphics[width=1\linewidth]{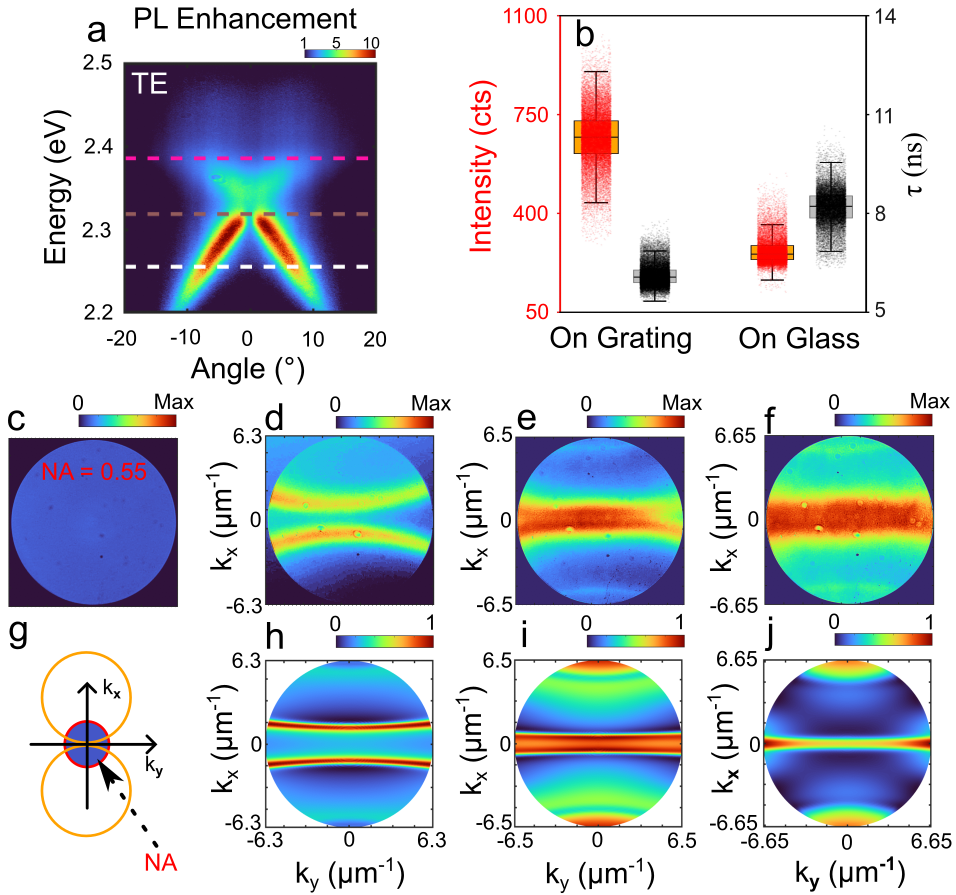}
\caption{\textbf{PL enhancement and BFP images.} a) The apparent PL enhancement in the map of energy and angles. It is determined by the normalization of Fig. 3b by Fig. 3a. b) The median PL intensity (in red) and PL lifetime (in black) of  CsPbBr$_3$ NCs on glass and on the TiO$_2$ grating for the scanned area (58 $\mu$m x 90 $\mu$m). c)-f) Experimental PL intensity at the back focal plane of the objective for c) CsPbBr$_3$ NC film on glass at E = 2.38 eV; and CsPbBr$_3$ NC film on the TiO$_2$ grating at d) E = 2.38 eV (pink dash line in a)), e) E = 2.32 eV (brown dash line in a)), and f) E = 2.25 eV (white dash line in a)). g) a sketch of the emission cones in momentum space for 1D grating. The orange circle indicates the diffraction cone at a given energy, and the blue-filled circle shows the radiated light cone captured by the numerical aperture (NA, red circle) of the objective. h-j) Simulated BFP images performed with Comsol Multiphysics for the corresponding energy values of d-f.}
\label{fig4}
\end{figure}

To quantify the coupling efficiency, we normalized the PL intensity of NCs on the TiO$_2$ grating (Fig. 3b) against that on glass (Fig. 3a) and plotted the results in Fig. 4a. The enhancement pattern closely reflects the measured optical modes (Fig. 2c), with the maximum 10-fold enhancement occurring near the BIC point, which decreases slightly as the emission angle increases. The enhancement of the bright mode is slightly smaller than that for the dark mode. The result of PL enhancement for TM mode is presented in Fig. S5b of the SI. To calculate the Purcell factor, we deployed a direct estimation of the Purcell factor based on average lifetime of the perovskite NCs on the TiO$_2$ grating and on glass. To do this, we computed the PL intensity and lifetime across the entire scanned area (58 $\mu$m x 90 $\mu$m) which includes the NCs on glass and TiO$_2$ grating. Figure 4b shows the median values of the PL intensity and lifetime of all the recorded wavelengths of NCs on the glass (256 cts) and TiO$_2$ grating (670 cts). The results show good agreement with those presented in Fig. 3h, provided by a larger sampling of the data. The mean PL lifetimes of the perovskite NCs on the TiO$_2$ grating for TE mode in the entire scanned area is $\tau_{grating}$ = 6.1 ns, compared to $\tau_{glass}$= 8.2 ns for the perovskite NC film on the glass. The Purcell factor for the examined area can be directly estimated FP=$\tau_{glass}$⁄$\tau_{grating}$ =1.34.77 The factor is consistent with other report,61 which validate the occurrence of the Purcell effect in the system. While other studies have noted PL enhancement of organic dyes coupled with dielectric (TiO$_2$ and Zr0$_2$) metasurfaces, the observed Purcell factor in those cases seems negligible.\cite{Murai2017,Higashino2022} This discrepancy may be due to the limited number of emitters exposed to the enhanced electric fields sustained by the dielectric metasurface and/or the overall quality of the emitters. A similar result for TM modes was also observed, with the PL enhancement following the features of Bloch modes, as illustrated in Fig. S7c of the SI. The fluorescence enhancement factor for TM mode reaches its maximum of 5-fold at angles around 15$^{\circ}$, being somewhat smaller compared to the TE mode case. The results revealed that the emission of the perovskite NCs has radically changed from isotropic, spatially incoherent, and unpolarized on the glass substrate to directional, spatially coherent, and polarized when coupled with the Bloch resonances supported by the TiO$_2$ grating. The coupling between  the Bloch resonances and the emission of the perovskite NC film led us to investigate the BFP image in (kx, ky) at isofrequencies marked by horizontal dash lines in Fig. 4a. The emission patterns of the perovskite NCs on glass substrate at the three frequencies are unpolarized and spatially incoherent, resulting in a homogeneous region in the BFP image, as illustrated in Fig. 4c. In contrast, the PL signal from the perovskite NCs on the TiO$_2$ grating exhibited distinct emission patterns at those isofrequencies. Specifically, BFP images of three iso frequencies on the TiO$_2$ grating centered at 2.38 eV (520 nm), 2.32 eV (535 nm), and 2.25 eV (550 nm), are displayed in Figs. 4d, 4e, and 4f, respectively. The PL signal is now concentrated in arcs, indicating that the light emission of the perovskite NCs, coupled with the Bloch resonances of the TiO$_2$ grating, has resulted in spatially coherent light and increased light extraction. To explain the arcs observed in BFP images, a sketch of optical modes at an isofrequency (represented  by orange circles) in momentum space is shown in Fig. 4g. The blue-filled circle represents the NA captured by the objective. Figures 4h-4j show the simulated BCF images at the associated frequencies of the dashed lines in Fig. 4a-4c. The BCF simulation is performed with Comsol Multiphysics via calculation of the reflectivity under simultaneous sweeping of both the incident angle (0$^{\circ}$ $\leq$ $\theta$ $\leq$ 34 $^{\circ}$) and the azimuthal angle of xy plane (0 $^{\circ}$ $\leq$ $\phi$ $\leq$ 90 $^{\circ}$). A good agreement was achieved between the measured and simulated BFP images, thus verifying the light outcoupling as well as its emissive directivity and polarization are shaped by the coupling with Bloch resonances supported by the TiO$_2$ grating.

In conclusion, we have explored the interaction between the 1D TiO$_2$ grating and the  CsPbBr$_3$ NC film deposited on top. Utilizing BFP spectroscopy and FLIM measurements, we observed significant modifications in PL intensity, directivity, polarization, and spectral shape of the perovskite NCs’ emission induced by the coupling with Bloch resonances supported by the TiO$_2$ grating. Through both simulation and measurement of angle-resolved reflectance, we confirmed the presence of the Bloch resonances and illustrated how their optical fields modified the local density of states surrounding the perovskite NCs. The BFP spectroscopy revealed a transformation of the unpolarized and spatially incoherent emission of perovskite NC film into a polarized and coherent one, driven by the optical resonances, thereby enhancing the light out- coupling. These findings contribute to further understanding of how nanophotonic structures can effectively control the emission properties and light extraction of the perovskite NCs films, which may pave the way for developing advanced colloidal nanocrystal-based optoelectronic devices. Future studies will be focused on pushing toward the mapping of light-matter interaction at nanoscale, quantifying the enhancement of the PLQY of NCs film deposited on optical metasurfaces, and optimizing the outcoupling efficiency of real optoelectronic devices.

\section{ASSOCIATED CONTENT}
\subsection{Supporting Information}

Synthesis of perovskite CsPbBr$_3$ NCs, characterizations of CsPbBr$_3$ NC film, fabrication procedure of TiO$_2$ grating, angle-resolved reflectivity and PL measurement setup, numerical simulations with Comsol Multiphysics, angle-solved PL results in TM mode, FLIM measurement and analysis.

\subsection{Acknowledgements}
This research was supported by Vingroup Innovation Foundation under contract number VinIF.2021.00169. A. Rogach acknowledges support from the Croucher Foundation of Hong Kong SAR, and from Qingdao Innovation and Development Center of Harbin Engineering University (P.R. China). S. T.  Ha, T. T. H. Do and V. Valuckas gratefully acknowledge the financial support from Singapore MTC-Programmatic Grant No. M21J9b0085. H. S. Nguyen acknowledges support by the Auvergne-Rhône-Alpes region in the framework of PAI2020.
%\bibliography{PeNCsGrating}

\onecolumngrid

\begin{center}
	\textbf{\large --- SUPPLEMENTAL MATERIAL ---}
\end{center}

\setcounter{equation}{0}
\setcounter{figure}{0}
\setcounter{table}{0}
\setcounter{page}{1}
\setcounter{section}{0}

\section{Perovskite NCs synthesis.}
\subsection{Chemicals}
1-Octadecene (ODE, technical grade, 90$\%$), oleic acid (OA, 90$\%$), Cesium Carbonate (Cs$_2$CO$_3$, 99$\%$), Lead acetate trihydrate (PbAc$_2$·3H$_2$0, 99.99$\%$), Strontium acetate (SrAc$_2$), Octane (anhydrous, 99$\%$), Sodium dodecylbenzenesulfonate (Na-DBSA, technical grade) and Toluene (anhydrous, 99.8$\%$) were purchased from Sigma-Aldrich. Isopropyl alcohol (IPA) and ethyl acetate (anhydrous, 99.5$\%$) were purchased from J$\&$K Scientific Ltd. Didodecylmethylamine (DiDodMA) and Bz-Br (98$\%$) were purchased from Tokyo Chemical Industry (TCI). All reagents were used as received without any further experimental purification.

\subsection{Perovskite quantum dots synthesis and film preparation}

\textbf{Synthesis of CsPbBr$_3$ NCs}. 38 mg of PbAc$_2$·3H$_2$0 (0.1 mmol), 16 mg of Cs2CO3 (0.05 mmol), 103 mg of SrAc$_2$ (0.5 mmol), 1.5 ml of OA and 6 ml of ODE were combined in a 25 ml 3-neck flask, which was placed into the heating mantle on top of a stirring plate. The vacuum was applied first at room temperature for 5 min as a pre-degassing step. Within a few minutes, the contents of the flask were heated to 110 $^{\circ}$C under stirring and were kept for approximately 60 minutes. After the metal salts had been degassed, the flask was switched from being under vacuum to being under nitrogen flow. The temperature of reaction mixture was raised to 180 °C under stirring until all the metal salts dissolved, yielding a clear and colorless solution without any visible solids inside. Next, the reaction mixture was allowed to cool and stabilize at 100 $^{\circ}$C. In the glovebox, 100 $\mu$l of Bz-Br was added to 1 ml of pre-degassed ODE in a 4 ml glass vial. Next, 150 $\mu$l of DiDodMA was added with 1350 $\mu$l of ODE in another 4 ml glass vial. The resulting mixture of ligands was shaken by hand, yielding a clear and colorless solution. The two bottles are sealed and then removed from the glove box. The Bz-Br and DiDodMA vials were taken outside the glovebox, and 550 $\mu$l of Bz-Br solution was injected into the flask using a disposable syringe equipped with a 21G needle at 100 $^{\circ}$C. Upon the injection of the Bz-Br precursor, the reaction mixture immediately turned light yellow (forming CsPbBr$_3$ seeds). After stirring for 1 min, 1 ml of as-prepared DiDodMA solution was injected. Upon the injection of the DiDodMA precursor, the reaction mixture immediately turned to green and turbid. After injection, the reaction was quenched by replacing the heating mantle with an ice-water bath. The reaction was left to cool until it reached 40 $^{\circ}$C, after which the water bath was removed. During both injection and cooling, the reaction was kept under a nitrogen atmosphere. Once the reaction had cooled down to near room temperature, the crude solution was centrifuged at 5000 rpm for 5 min. After the centrifuge, the green supernatant was discarded. Then, 4 ml of anhydrous toluene was added to the vial to disperse all the precipitated polyhedral NCs. 5 $\mu$l of OA and 10 $\mu$L of Na-DBSA (5 mg/mL, dissolved in a mixture of toluene and IPA) was added for ligand exchange. ethyl acetate was then added with equal volume. The mixture was centrifuged at 7800 rpm for 5 min and redispersed into 0.5 mL of octane. The NC solution was centrifuged at 7800 rpm for 5 min, yielding very limited yellow precipitate on the walls of the vial. The CsPbBr$_3$ NC film was spin coated on the sample at a speed of 1000 rpm for 45 s, followed by baking the film at 80 $^{\circ}$C for 5 min. The perovskite NC films were stored in ambient conditions.

\subsection{Characterizations of the NCs film}
XRD pattern of CsPbBr$_3$ NC film deposited on a Si substrate was recorded using a Bruker X-Pert X-ray diffractometer using Cu K$\alpha$ radiation ($\lambda$ = 1.5418 $\mathring{A}$). The high-resolution TEM was performed using Jeol 2100F machine.
Absorption of the NC films was measured with a UV- Vis spectrometer (Colo Novel-1025) and PL was acquired by using the PL setup, see Fig. S2, under CW laser $\lambda$  = 405 nm with a low excitation power. We measured the PLQY of the CsPbBr$_3$ solid after a week of exposure to the humid air (humidity was approx. 70$\%$). The PLQY was measured using a calibrated integrating sphere in Edinburgh Instruments model FLS1000 spectrofluorometer. Both excitation and emission paths are coupled with grating monochromators, the light source is a Xenon lamp, while the exit of the emission monochromator is a photomultiplier tube (PMT) detector.

\begin{figure}[ht!]
\centering
\includegraphics[width=0.8\linewidth]{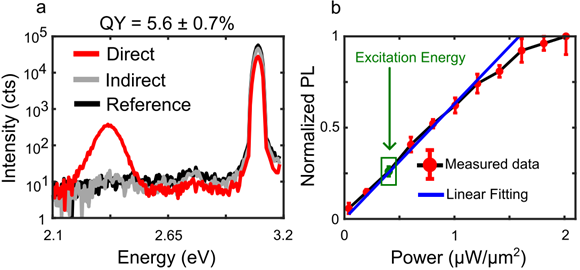}
\caption{\textbf{PLQY measurement of CsPbBr$_3$ NC film with three-measurement technique.} Direct measurement stands for the measurement under direct laser excitation on the CsPbBr$_3$ NC film, indirect one refers to the measurement in which the laser beam is hit the sphere’s wall first and only scattered light is absorbed by the NC film, and reference measurement was performed in the indirect configuration with blank quartz substrate in the sample position. b) The dependance of PL intensity on the power of laser. Green box indicate for pumping power value used in the paper which into the linear respond region.}
\label{figS1}
\end{figure}

PL spectra were collected under excitation of 3.07 eV (403.8 nm) from a broadband Xenon light source. Five different positions on the film were recorded to yield the average PLQY is 5.6$\%$. Integrated emission signal for calculating PLQY was narrowed from 2.17 eV to 2.53 eV, as shown in Fig. S1.

\section{Fabrication of TiO$_2$ grating nanostructures}
A layer of TiO$_2$ film (120 nm thick) was deposited by ion-assisted sputtering (Oxford Opto- fab3000) on a glass substrate. A 30 nm-thick Cr film was deposited directly on the TiO$_2$ by electron beam evaporation (Angstrom EvoVac). A layer of negative electron beam resist Hydrogen Silsesquioxane (HSQ, Dow Corning) was spin coated on top of Cr at 5000 rpm for 60 seconds followed by sequential baked at 120 $^{\circ}$C for 2 min and 180 $^{\circ}$C for another 2 min. The resist was then exposed to an electron beam (accelerated voltage = 100 keV and current = 500 pA) in the Elionix ELS-7000 system. The HSQ pattern was then developed in the salty developer solution consisting of 1 wt.$\%$ NaOH and 4 wt.$\%$ NaCl in deionized water. The pattern was transferred onto Cr layer by using inductively coupled plasma reactive ion etching with a mixture of Cl2:O2 gas (19 sccm:2 sccm) and subsequently transferred onto the TiO$_2$ layer by using CHF3 gas. All the etching of Cr and TiO$_2$ was performed with Oxford PlasmaPro 100 Cobra device. Finally, the layer of Cr and HSQ masks were removed by immersing the sample in a Cr etchant solution. The fabricated nanostructures were examined by a scanning electron microscope (Hitachi SU8220).

\section{Optical measurements}

PL spectra, angle-resolved reflectivity, and angle-resolved PL were acquired using a Princeton Instruments monochromator and s-CMOS camera, Kuro Model 1200. The laser source used for PL measurements is a 405 nm CW laser with a low excitation energy is 0.4$\mu$W/$\mu$ m$^{2}$  to ensure emission intensity is in the linear range with the excitation power. When switching to angle-resolved reflectivity, the laser is replaced by the broadband lamp (Thorlabs SLS202). In angle-resolved reflectivity, we deployed the cross-polarization measurement to probe the purely guided resonances in the system. Two pairs of linear polarizers and half-wave plates: one at the incident beam and the other at the output path (after the beam splitter) are used to filter reflected background signals resulting from reflections of the multiple interfaces between material layers. The back-focal imaging was achieved by inserting a band-pass filter prior to the pinhole to select the interested wavelengths and the images are projected on a CMOS camera.

\begin{figure}[ht!]
\centering
\includegraphics[width=1\linewidth]{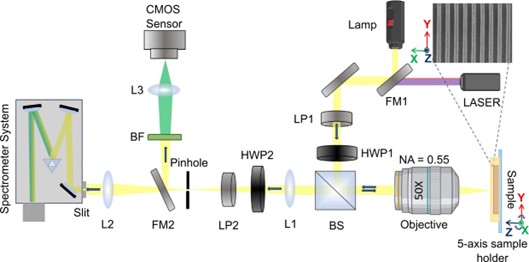}
\caption{\textbf{A diagram of the PL and reflectivity setups.} In PL mode, a continuous laser emitting at 405 nm was used, whereas the broadband lamp was utilized for reflectivity measurement. The sample was placed on a 5-axis sample holder to obtain high-precision measurement. The TiO2 grating was placed so that its long groove direction is perpendicular to the vertical slit of the monochromator (Princeton Instrument HRS-300). In PL measurement, the laser beam is focused on the sample by a long working distance objective (NA = 0.55, PAL-50, OptoSigma). The emissive light at the back-focal plane of the objective is passed through a beam splitter (BS), and a Fourier lens (L1, f = 150mm, LA1222-AB, Thorlabs). A long-pass filter ($\lambda_{cut-off}$  = 495 nm, FGL495M, Thorlabs) is placed after Fourier lens (L1) to filter out the excitation laser beam, and an analyzer set including a linear polarizer (LP1\&2, DTM-SPF-30C-32, OptoSigma) and a half-wave plate (HWP1\&2, AHWP10M-600, Thorlabs) is used to analyze the emissive light. A flipping mirror (FM2) is , MHGT-25.4, OptoSigma) switched on and off the optical path to change between the spectroscopy and microscopy lines. Two lenses L2 (f = 150mm, LA1222-AB, Thorlabs) and L3 (f = 150mm, LA1222-AB, Thorlabs) are used to focus the signal on the spectrometer (Princeton Instrument Kuro 1200) and CMOS camera, respectively. In the back focal plane imaging measurement in Fig. 4, a bandpass filter (FBH560, Thorlabs) was used to select the right wavelengths with the angle of incidence light finely tuned to gain the desired wavelengths. }
\label{figS2}
\end{figure}

\section{Numerical simulations with COMSOL Multiphysics}

This section describes the simulated results of Figs. 2e, 2f, and Fig. 4h-4j in the main text. The angle-resolved reflectivity (used for Figs. 2e and 2f) was simulated using the wave optics module of Comsol Multiphysics. Floquet boundary conditions were applied to each unit cell of the grating to simulate an infinite grating. The excitation port was placed above the perovskite solid to resemble the experimental configuration, and the first order of diffraction was added to the excitation port. The refractive indices of glass and TiO2 are nglass = 1.46 and nTiO2 = 2.54, respectively. The dimensions of the TiO2 stripes were imported from the SEM image. In the simulation, the period of the unit cell is p = 260 nm and width = 158 nm in the x-direction and the thickness of the perovskite film is 400 nm with an effective relative permittivity fitted with the Lorentz oscillator model.
\begin{equation}\label{eq:relative_permit}
\epsilon(E)= n^{2}+\frac{A_{X}}{E_{0X}^{2}-E^2-i\gamma_{X}E}
\end{equation}

where $n = 1.69$ is effective the refractive index of the passive structure, $A_{X} =$ 0.167 eV$^{2}$ is the oscillator strength exciton, $E_{0X} =$ 2.39 eV and $\gamma_{X} =$ 68  meV.

\begin{figure}[ht!]
\centering
\includegraphics[width=0.4\linewidth]{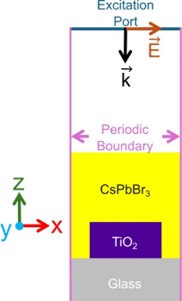}
\caption{\textbf{A sketch of simulation model in Comsol Multiphysics.} The angle-resolved reflectivity was performed by varying the incident angle theta (0$^{\circ}$ $\leq$ $\theta$ $\leq$ 34 $^{\circ}$) from the incident port.}
\label{figS3}
\end{figure}

The simulation of the structure near the excitation energy is presented in Fig. S4. It is important to note that the real component of the permittivity ($n^{2} = 1.69^{2}$) was kept unchanged for the absorption simulation, while the imaginary part was taken from Ref [S1] \cite{Brennan2024}. Figure S4a illustrates that no noticeable optical mode exists at the excitation energy; however, the presence of the TiO$_2$ grating might enhance the excitation light scattering. Thus, the absorption of the perovskite NC layer on the grating and on the glass at the excitation energy (E = 3.06 eV) was plotted as a function of the excitation angle in Fig. S4b. A slight increase in the absorption of the perovskite layer positioned on the grating can be observed compared to that on the glass. The excitation enhancement across all excited angles was calculated as:

\begin{equation}\label{eq:equation_1}
\Lambda(exc)={\int_{0}^{\phi} A_{grating}(\phi)\,d\phi} / { \int_{0}^{\phi} A_{glass}(\phi)\,d\phi }=1.15
\end{equation}

where $\phi$ represents maximal angle of the objective, and $A(\phi)$ denotes the absorptance of the perovskite NCs layer on the TiO$_2$ grating and on the glass substrate. Although the enhancement at excitation energy appears to be modest ($\Lambda(exc)$=1.15), it is important to note that if the pure pumping enhancement on the ensembles of the NCs occurs, f pure pumping enhancement were to occur in the ensembles of the perovskite NCs, it would result in an increase in their PL intensity without altering the spectral profile. However, our experimental results indicate that the spectral profile of the perovskite layer on the TiO$_2$ grating was significantly influenced by the modes of the Bloch resonances. Therefore, we conclude that the impact of excitation on the interaction between the TiO$_2$ grating and the perovskite NC film can be considered negligible.
 To simulate the back focal images presented in Figs. 4h-4i, we still use the above model except that the excitation energy was fixed at constant values, E= 2.38 eV (Fig. 4h), E= 2.32 eV (Fig. 4i), and E= 2.25 eV (Fig. 4j). The incident angle ($\theta$) and the azimuthal angle ($\phi$) were swept simultaneously. We extracted the reflectance and presented it as the back focal images.

\begin{figure}[ht!]
\centering
\includegraphics[width=1\linewidth]{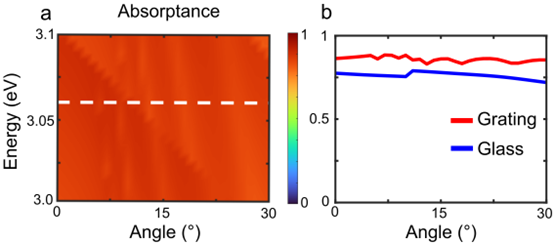}
\caption{a) Absorption spectra of the TiO$_2$ grating covered by a CsPbBr$_3$ NC film around excitation wavelength (marked by horizontal white dash line). No optical mode is visible in this region. b) A reflectance spectrum at the excitation energy of the laser as a function of angle. Due to the scattering improvement, we calculated the increased factor of absorption on the perovskite layer in the presence of the grating is 1.15.}
\label{figS4}
\end{figure}

\section{Angle-resolved PL in TM mode}
Figure S5a shows the TM mode for angle-resolved PL of the perovskite NCs film deposited on top of the grating. The PL intensity distribution differs from that of TE mode (Fig. 3b in the main text). A slight decrease in the intensity and the enhancement value occurred at a higher angles, compared to TE mode, can be observed.

\begin{figure}[ht!]
\centering
\includegraphics[width=1\linewidth]{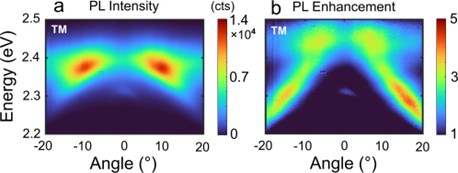}
\caption{a) Angle-resolved PL measurement result of the CsPbBr$_3$ NC film on the TiO$_2$ grating in TM mode. b) The apparent PL enhancement in the map of energy and angles.}
\label{figS5}
\end{figure}

We also normalized the PL of NCs on the grating with the PL of NCs film on the glass. Figure S5b displayed the apparent PL with a five-fold enhancement observed at 2.3 eV and angle $\phi$ = 17$^{\circ}$, which is two times smaller than the enhancement observed in TE mode in Fig. 4b of the main text. Importantly, the PL enhancement in TM mode follows exactly the features of the reflectance in TM mode presented in Fig. 2c in the main text.

\section{FLIM measurements}
A pulsed diode laser source ($\lambda$ = 405 nm, repetition rate = 10 MHz) with adjustable intensity was used to excite the system through a polarization-maintained fiber. The excitation beam passes through a series of mirrors and a dichroic beam splitter (BS), and then reaches the motorized and synchronized galvo scanning mirror. The scanning mirror steers the excitation beam to lateral positions on the sample, which is coupled with an inverted Nikon microscope (Ti2-U) and a 20X objective (NA = 0.75). The emitted light is filtered by the dichroic beam splitter ($\lambda_{cut-off}$=425 nm), passes to a linear analyzer and focused on the detector (GaAsP PMT). A complete description of the FLIM setup was discussed in Ref [xx] of the main text. The signal is finally stored in a metafile and further processed by VistaVision software using a single exponential fitting. 

\begin{figure}[ht!]
\centering
\includegraphics[width=1\linewidth]{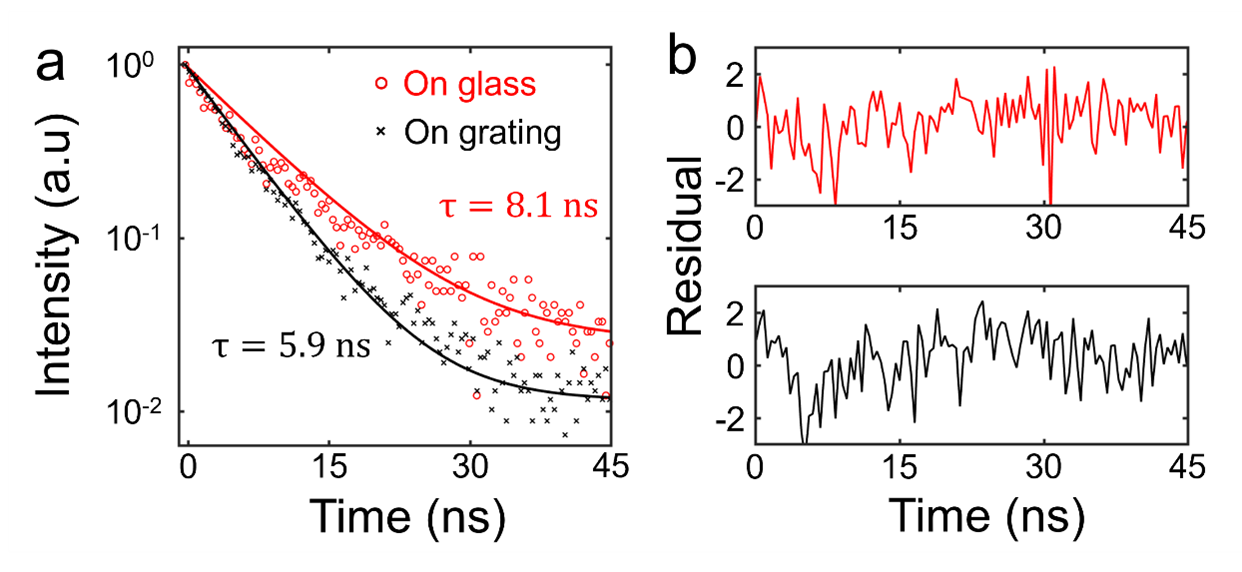}
\caption{a) The PL intensity decay fittings for two random points: one on the glass at coordinate (x$_{1}$ = 15 $\mu$m, y$_{1}$ = 25 $\mu$m) and one on the TiO$_{2}$ grating at (x$_{2}$ = 48 $\mu$m, y$_{2}$ = 25 $\mu$m). b) The residuals of the lifetime fittings of NCs are located on the glass (red) and on the TiO$_{2}$ grating (black).}
\label{figS6}
\end{figure}

We plotted the PL lifetimes (obtained from a single exponential fitting) of two representative points collected on the glass and on the TiO$_{2}$ grating in Fig. S6a. The PL lifetime of perovskite NCs on the TiO$_{2}$ grating is reduced to 5.9 ns, compared to 8.1 ns for the NC film on the glass.  The quality of fitting is shown to be good with over 99$\%$ of the residual values being between -2 and 2 (Fig. S6b). 

\begin{figure}
\centering
\includegraphics[width=0.8\linewidth]{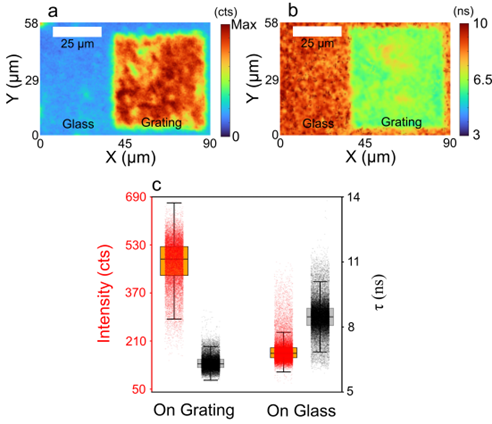}
\caption{a) PL intensity distribution of the CsPbBr$_{3}$ NC film on the glass and on the TiO$_{2}$ grating in TM mode with FLIM measurement. b) PL lifetime extracted from a) with a single exponential fitting. c) Statistical analysis for FLIM measurements. The median values of PL intensity and lifetime of the CsPbBr$_{3}$ NCs on glass and on the TiO$_{2}$ grating for TM mode.}
\label{figS6}
\end{figure}

From the single exponential decay fitting, we obtained the corresponding lifetime mapping for the PL mapping of the TM mode (presented in Fig. S7a-b). The pixel-based analysis of the PL intensity and corresponding lifetimes in TM modes is presented in Fig. S7c. It provides insight into the correlation between PL intensity and the lifetime of the perovskite NCs deposited on the TiO2 grating and on the glass substrate. The results demonstrate the inverse relationship between emission intensity and lifetime; as the emission intensity increases, the corresponding lifetime decreases. In the TM mode, the median PL emission intensity of perovskite NCs on glass is 169 cts, while on the grating it is significantly higher and reaches the value of 483 cts. Conversely, the median lifetime of NCs coupled with the grating is 6.3 ns, notably shorter than the 8.5 ns lifetime observed for perovskite NCs on glass. 

\bibliography{PeNCsGrating}

\end{document}